\documentclass[12pt]{article}

\usepackage{epsf}
\usepackage{latexsym,amssymb,euscript}
\usepackage[dvips]{graphicx}
\usepackage{amsmath}

\begin{document}

\begin{center}
{\Large \bf  Phase structure of $3D$ $Z(N)$ lattice gauge theories at 
finite temperature:  large-$N$ and continuum limits }
\end{center}

\vskip 0.3cm
\centerline{O.~Borisenko$^{1\dagger}$, V.~Chelnokov$^{1*}$, 
M.~Gravina$^{2\ddagger}$, A.~Papa$^{2\P}$}

\vskip 0.6cm

\centerline{${}^1$ \sl Bogolyubov Institute for Theoretical Physics,}
\centerline{\sl National Academy of Sciences of Ukraine,}
\centerline{\sl 03680 Kiev, Ukraine}

\vskip 0.2cm

\centerline{${}^2$ \sl Dipartimento di Fisica, Universit\`a della 
Calabria,}
\centerline{\sl and Istituto Nazionale di Fisica Nucleare, 
Gruppo collegato di Cosenza}
\centerline{\sl I-87036 Arcavacata di Rende, Cosenza, Italy}

\vskip 0.6cm

\begin{abstract}
We study numerically three-dimensional $Z(N)$ lattice gauge theories 
at finite temperature, for $N = 5,\ 6,\ 8,\ 12,\ 13$ and 20 on lattices with 
temporal extension $N_t=2,\ 4,\ 8$. For each model, we locate phase transition 
points and determine critical indices. We propose also the scaling of 
critical points with $N$. The data obtained enable us to verify the scaling 
near the continuum limit for the $Z(N)$ models at finite temperatures. 
\end{abstract}

\vfill
\hrule
\vspace{0.3cm}
{\it e-mail addresses}:
$^\dagger$oleg@bitp.kiev.ua, \  $^*$chelnokov@bitp.kiev.ua,
\  $^{\ddagger}$gravina@fis.unical.it, \ \ $^{\P}$papa@fis.unical.it 

\newpage 

\section{Introduction} 

The deconfinement phase transition in finite-temperature lattice gauge 
theories (LGTs) has been one of the main subjects of investigation for the 
last three decades. In this paper we concentrate on $Z(N)$ LGTs, which 
are interesting on their own and can provide for useful insights into the 
universal properties of $SU(N)$ LGTs, being $Z(N)$ the center subgroup of 
$SU(N)$. 

The most general action for the $Z(N)$ LGT can be written as 
\begin{equation} 
S_{\rm gauge} \ = \ \sum_x \sum_{n<m} \ \sum_{k=1}^N \beta_k
\cos \left( \frac{2 \pi k}{N} \left(s_n(x) + s_m(x+e_n) 
-s_n(x+e_m) - s_m(x) \right) \right) \ .
\label{action_gauge}
\end{equation}
Gauge fields are defined on links of the lattice and take on values 
$s_n(x)=0,1,\cdots,N-1$. $Z(N)$ gauge models, similarly to their spin cousins, 
can generally be divided into two classes - the {\em standard Potts} models
and the {\em vector} models. 
The standard gauge Potts model corresponds to the choice when all 
$\beta_k$ are equal. Then, the sum over $k$ in~(\ref{action_gauge}) reduces 
to a delta-function on the $Z(N)$ group. The conventional vector model 
corresponds to $\beta_k=0$ for all $k>1$. For $N=2,\ 3$ the Potts and vector 
models are equivalent. 

For an extended description of the phase structure of $Z(N)$ LGTs
in three dimension and for a list of references, we refer the reader 
to our recent papers~\cite{3d_zn_strcoupl,ZN_fin_T,ZN_zero_T}, where 
also a detailed description of our motivations can be found. 

In those papers we have initiated exploring the phase structure of the vector 
$Z(N)$ LGTs for $N>4$. More precisely, we have first considered an 
anisotropic lattice in the limit where the spatial coupling 
vanishes~\cite{3d_zn_strcoupl}. In this limit the spatial gauge fields can be 
exactly integrated out and one gets a $2D$ generalized $Z(N)$ model. The 
Polyakov loops play the role of $Z(N)$ spins in this model. For the Villain 
version of the resulting  model we have been able to present 
renormalization group (RG) arguments indicating the existence of two BKT-like 
phase transitions:

- a {\em first transition}, from a symmetric, confining phase to
an intermediate phase, where the $Z(N)$ symmetry is enhanced to $U(1)$ 
symmetry;

- a {\em second transition}, from the intermediate phase to a phase with
broken $Z(N)$ symmetry. 

This scenario was confirmed with the help of large-scale 
Monte Carlo simulations of the effective model. We have also computed some 
critical indices, which appear to agree with the corresponding indices of $2D$ 
$Z(N)$ spin models, thus giving further support to the Svetitsky-Yaffe 
conjecture~\cite{Svetitsky}. In particular, we found that the magnetic
critical index $\eta$ at the first transition, $\eta^{(1)}$, takes the
value 1/4 as in $2D$ $XY$, while its value at the second transition, 
$\eta^{(2)}$, is equal to $4/N^2$.

Then, we extended our analysis to the full isotropic $3D$ $Z(N)$ LGT  
at finite temperature~\cite{ZN_fin_T}. It is well known that the full phase 
structure of a finite-temperature LGT is correctly reproduced in the limit 
where spatial plaquettes are neglected. They have probably small influence 
on the dynamics of the Polyakov loop interaction. Therefore,  
the scenario advocated by us in~\cite{3d_zn_strcoupl} was expected to 
remain qualitatively correct for the full theory. It was indeed confirmed
by numerical Monte Carlo simulations~\cite{ZN_fin_T} that the full gauge 
models with $N>4$ possess two phase transitions of the BKT type, with 
critical indices coinciding with those of $2D$ vector spin models. 

The aim of the present paper is to 

\begin{itemize} 
\item 
extend the study of Ref.~\cite{ZN_fin_T} to other values of $N$ and to 
$N_t=8$;

\item 
check the scaling near the continuum limit and establish the scaling formula 
for critical points with $N$. 
\end{itemize}

In particular the study of the continuum limit in this work is an important 
step forward with respect to Ref.~\cite{ZN_fin_T}. The theory of dimensional 
cross-over~\cite{caselle} explains how critical couplings and indices of a 
finite temperature LGT (finite $N_t$) approach critical couplings and indices 
of the corresponding zero-temperature theory ($N_t \to \infty$). This provides
us with a way to crosscheck our zero-temperature results~\cite{ZN_zero_T}
and thus predict the critical temperature in the continuum limit.

The BKT transition is hard to study analytically, by, say, the RG of 
Ref.~\cite{RG}. On the other side, numerical simulations are plagued by very 
slow, logarithmic convergence to the thermodynamic limit near the transition,
thus calling for large-scale simulations, together with finite-size scaling
(FSS) methods. The standard approach would consist in using Binder cumulants 
and susceptibilities of the Polyakov loop to determine critical couplings and 
critical indices. Here, as in Ref.~\cite{ZN_fin_T}, we follow a different
strategy: we move to a dual formulation and use Binder cumulants and 
susceptibilities of {\em dual $Z(N)$ spins}. This has some important
consequences: (i) the critical behavior of dual spins is reversed with respect 
to that of Polyakov loops, namely the spontaneously-broken ordered phase is
mapped to the symmetric phase and {\it vice versa}; (ii) the magnetic 
critical indices $\eta$ are interchanged, whereas the index $\nu$ is expected 
to be the same (=1/2) at both transitions (see Ref.~\cite{ZN_fin_T} for 
details). The obvious advantage of this approach is that cluster algorithms 
become available, with considerable speed up in the numerical procedure. 

This paper is organized as follows: in Section~2 we recall the connection of 
our model with a generalized $3D$ $Z(N)$ spin model; in Section~3 we present 
the setup of Monte Carlo simulations and our numerical results 
for critical points and critical indices; in the same section we study
also the scaling with $N$ of critical couplings and the continuum limit;
finally, in Section~4 we discuss our results and the open problems.

\section{Theoretical setup}

Duality amounts to map a theory based on gauge links to a spin theory
and, therefore, it opens the doors to Monte Carlo simulations by cluster 
algorithms, which make the spin theory much easier to be studied by numerical 
methods. In this work, following the strategy of Ref.~\cite{ZN_fin_T},
we study the phase structure of the $3D$ LGT defined in~(\ref{action_gauge})
simulating its dual $3D$ $Z(N)$ spin model. We briefly recall here the
main issues related with the duality transformation.

The $3D$ $Z(N)$ gauge theory on an anisotropic $3D$ lattice $\Lambda$ can 
generally be defined as 
\begin{equation}
Z(\Lambda ;\beta_t,\beta_s;N) \ = \  \prod_{l\in \Lambda}
\left ( \frac{1}{N} \sum_{s(l)=0}^{N-1} \right ) \ \prod_{p_s} Q(s(p_s)) \
\prod_{p_t} Q(s(p_t)) \; ,
\label{PTdef}
\end{equation}
where the link angles $s(l)$ are combined into the conventional plaquette angle
\begin{equation}
s(p) \ = \ s_n(x) + s_m(x+e_n) - s_n(x+e_m) - s_m(x) \ .
\label{plaqangle}
\end{equation}
Here, $e_n$  ($n=0,1,2$) denotes a unit vector in the $n$-th direction and
the notation $p_t$ ($p_s$) stands for the temporal (spatial) plaquettes. 
Periodic boundary conditions (BC) on gauge fields are imposed in all 
directions. The most general $Z(N)$-invariant Boltzmann weight with $N-1$ 
different couplings is
\begin{equation}
Q(s) \ = \
\exp \left [ \sum_{k=1}^{N-1} \beta_p(k) \cos\frac{2\pi k}{N}s \right ] \ .
\label{Qpgen}
\end{equation}
The Wilson action corresponds to the choice $\beta_p(1)=\beta_p$, 
$\beta_p(k)=0, k=2,...,N-1$, which is the one adopted in this work. 
Furthermore, we will consider an isotropic lattice: $\beta_s=\beta_t=\beta$.

Our study is based on the mapping of the gauge model to a generalized $3D$ 
$Z(N)$ spin model on a dual lattice $\Lambda_d$, whose action is
\begin{equation}
\label{modaction}
S \ =\ \sum_{x}\ \sum_{n=1}^3 \sum_{k = 1}^{N-1} \ \beta_k \  
\cos \left( \frac{2 \pi k}{N} \left(s(x) - s(x+e_n) \right) \right) \ .
\end{equation}
The dual mapping is realized once one specifies the relationship between the 
original gauge coupling $\beta$ and the dual effective couplings $\beta_k$.
This has been done in Ref.~\cite{ZN_fin_T} (see also Ref.~\cite{ukawa}) and
the result is
\begin{equation}
\beta_k \ =\ \frac{1}{N} \sum_{p = 0}^{N - 1} \ln \left [ \frac{Q_d(p)}
{Q_d(0)} \right ] \  \cos \left(\frac{2 \pi p k}{N} \right) \ .
\label{couplings}
\end{equation}

In~\cite{ZN_fin_T} the explicit form was given for the connection between the
coupling $\beta$ of the LGT and the couplings $\beta_k$ of the dual spin model
in the case of $N=5$. Two features clearly emerged there: first, $\beta_1$ 
turned to be much larger (in absolute value) than $\beta_2$, thus suggesting 
that the $3D$ vector spin model with only $\beta_1$ non-vanishing gives 
already a reasonable approximation of the gauge model (in our simulations we 
use all $\beta_k$); second, the weak and the strong coupling regimes are 
interchanged, {\it i.e.} when $\beta\to\infty$ the effective couplings 
$\beta_k\to 0$ and, therefore, the ordered symmetry-broken phase is mapped to 
a symmetric phase with vanishing magnetization of dual spins. Conversely, the 
symmetric phase at small $\beta$ becomes an ordered phase where the dual 
magnetization is non-zero.
It turns out that the interchange of phases under the dual mapping is not a 
special feature of $N=5$, but is rather a general property valid for any $N$. 

In Ref.~\cite{ZN_fin_T} also an interesting phenomenon was discussed: at the
critical point $\beta_{\rm c}^{(1)}$, corresponding to the first transition
of the LGT (from the symmetric to the intermediate phase), the {\em dual} 
correlation function scales with a critical index $\eta$ equal to the
index $\eta^{(2)}=4/N^2$ of the Polyakov loop correlator in the LGT, while
at the critical point $\beta_{\rm c}^{(2)}$ of the second transition
in the LGT (from the intermediate to the broken phase), it scales with 
a critical index $\eta$ equal to the index $\eta^{(1)}=1/4$ of the Polyakov 
loop correlator in the LGT. This can be proved in the Villain formulation of
the $2D$ theory and only conjectured (but confirmed numerically) in the $3D$ 
case~\cite{ZN_fin_T}.

\section{Numerical setup and results}

The $3D$ $Z(N)$ spin model, dual of the $3D$ $Z(N)$ Wilson LGT, has been
simulated by means of a cluster algorithm on $N_t \times L \times L$ lattices 
with periodic BC. The system has been studied for $N = 5,\ 6,\ 8,\ 12,\ 13$
and 20 on lattices with the temporal extension $N_t=2,\ 4,\ 8$. With respect 
to our previous work~\cite{ZN_fin_T}, we considered new values of 
$N$ (6, 8, 12, 20) and included also $N_t=8$.

We focused on the following observables:

\begin{itemize}
\item complex magnetization $M_L = |M_L| e^{i \psi}$,
\begin{equation}
\label{complex_magnetization}
M_L \ =\  \sum_{x \in \Lambda} \exp \left( \frac{2 \pi i}{N} s(x) \right) \;,
\end{equation}
where we stress that $s(x)$ is a dual spin variable;

\item real part of the rotated magnetization, $M_R = |M_L| \cos(N \psi)$,
and normalized rotated magnetization, $m_\psi = \cos(N \psi)$;

\item susceptibilities of $M_L$ and $M_R$:  
$\chi_L^{(M)}$, $\chi_L^{(M_R)}$
\begin{equation}
\label{susceptibilities}
\chi_L^{(\mathbf\cdot)} \ =\  L^2 N_t \left(\left< \mathbf\cdot^2 \right> 
- \left< \mathbf\cdot \right>^2 \right)\;;
\end{equation}

\item Binder cumulants $U_L^{(M)}$ and $B_4^{(M_R)}$,
\begin{eqnarray}
U_L^{(M)}&\ =\ &1 - \frac{\left\langle \left| M_L \right| ^ 4 
\right\rangle}{3 \left\langle \left| M_L \right| ^ 2 \right\rangle^2}\;, 
\nonumber \\
\label{binderU}
B_4^{(M_R)}&\ =\ & \frac{\left\langle \left| M_R 
- \left\langle M_R \right\rangle \right| ^ 4 \right\rangle}
{\left\langle \left| M_R - \left\langle M_R \right\rangle \right| ^ 2 
\right\rangle ^ 2 } \ . 
\label{binderBMR}
\end{eqnarray}

\end{itemize}

\subsection{Critical couplings}

Studying numerically the behavior of the Binder cumulants $U_L^{(M)}$ and 
$B_4^{(M_R)}$ and the normalized rotated magnetization $m_\psi$ for different 
values of the lattice size $L$, we have determined the critical points using 
the following methods:

\begin{itemize}
\item as the {\em second transition point} $\beta_{\rm c}^{(2)}$, we have 
looked for the value of $\beta$ at which the curves giving the Binder cumulant 
$U_L^{(M)}(\beta)$ on lattices with different size $L$ ``intersect''. To be 
able to use different values of $L$, we defined the ``intersection point'' as 
the $\beta$ value at which the sum of the quadratic difference between all
possible pairs of values of $U_L^{(M)}$ is minimal over the chosen range of 
$L$ values ($192 \leqslant L \leqslant 768$). To improve the precision of the 
final result, following Ref.~\cite{3dxy_univ}, we Taylor-expanded the Binder 
cumulant up to the third order around $\beta = \beta_{\rm f}$, getting the 
coefficients of the expansion by the numerical simulation at $\beta_{\rm f}$, 
and repeated this procedure several times, each time taking $\beta_{\rm f}$ 
equal to the previous estimation of $\beta_{\rm c}$, making sure that these 
values do converge with iterations. 
The error bands on $\beta_{\rm c}^{(2)}$ were estimated as the largest among
the following differences between estimates of $\beta_{\rm c}$: 
a) difference between two consecutive iterations, 
b) difference between the estimates using $192 \leqslant L \leqslant 768$ 
and $192 \leqslant L \leqslant 512$, and 
c) difference between the estimates using $192 \leqslant L \leqslant 768$ and 
$256 \leqslant L \leqslant 768$. In most cases the third difference was the 
largest one.

\item The same method can in principle be used for the {\em first transition} 
$\beta_{\rm c}^{(1)}$ using either the Binder cumulant $B_4^{(M_R)}$ or 
$m_\psi$; it turned out, however, that the precision required by this 
method on these observables could not be met with a sensible simulation time.
For this reason, as the position of the first critical point we used our 
previous determinations given in Ref.~\cite{ZN_fin_T}, where 
$\beta_{\rm c}^{(1)}$ was estimated as the value of $\beta$ at which 
$B_4^{(M_R)}$ and  $m_\psi$ plotted {\it versus} $(\beta-\beta_{\rm c}^{(1)}) 
{\ln L}^{1/\nu}$ show the best overlap for different values of $L$. 
\end{itemize}

The results of the determinations of $\beta_{\rm c}^{(1)}$ and 
$\beta_{\rm c}^{(2)}$ are summarized in Table~\ref{tbl:crit_betas}.

\begin{table}[tb]
\caption{Values of $\beta_{\rm c}^{(1)}$ and $\beta_{\rm c}^{(2)}$ obtained 
for various $N_t$ in $3D$ $Z(N)$ with $N = 5,\ 6,\ 8,\ 12,\ 13\ \text{and}\ 
20$.}
\begin{center}
\begin{tabular}{|c|c|c|c|}
\hline
$N$ & $N_t$ & $\beta_{\rm c}^{(1)}$ & $\beta_{\rm c}^{(2)}$ \\
\hline
 5 & 2 & 1.617(2) & 1.6972(14) \\ 
 5 & 4 & 1.943(2) & 1.9885(15) \\ 
 5 & 6 & 2.05(1)  & 2.08(1) \\
 5 & 8 & 2.085(2) & 2.1207(9) \\ 
 5 & 12 & 2.14(1) & 2.16(1) \\
\hline
 6 & 2 & -        & 2.3410(15) \\ %
 6 & 4 & -        & 2.725(12) \\ %
 6 & 8 & -        & 2.899(4) \\ %
\hline
 8 & 2 & -        & 3.8640(10)\\ %
 8 & 4 & 2.544(8) & 4.6864(15) \\ 
 8 & 8 & 3.422(9) & 4.9808(5) \\ 
\hline
\end{tabular}
\hspace{2cm}
\begin{tabular}{|c|c|c|c|}
\hline
$N$ & $N_t$ & $\beta_{\rm c}^{(1)}$ & $\beta_{\rm c}^{(2)}$ \\
\hline
12 & 2 & -        & 8.3745(5) \\ %
12 & 4 & -        & 10.240(7) \\ %
12 & 8 & -        & 10.898(5) \\ %
\hline
 13 & 2 & 1.795(4) & 9.735(4) \\ 
 13 & 4 & 2.74(5)  & 11.959(6) \\ 
 13 & 8 & 3.358(7) & 12.730(2) \\ 
\hline
 20 & 2 & -       & 22.87(4)   \\ %
 20 & 4 & 2.57(1) & 28.089(3) \\ 
 20 & 8 & 3.42(5) & 29.758(6)   \\ 
\hline
\end{tabular}
\end{center}
\label{tbl:crit_betas}
\end{table}

\subsection{Scaling of the critical coupling with $N$}

\begin{figure}[tb]
\centering
\includegraphics[width=0.32\textwidth]{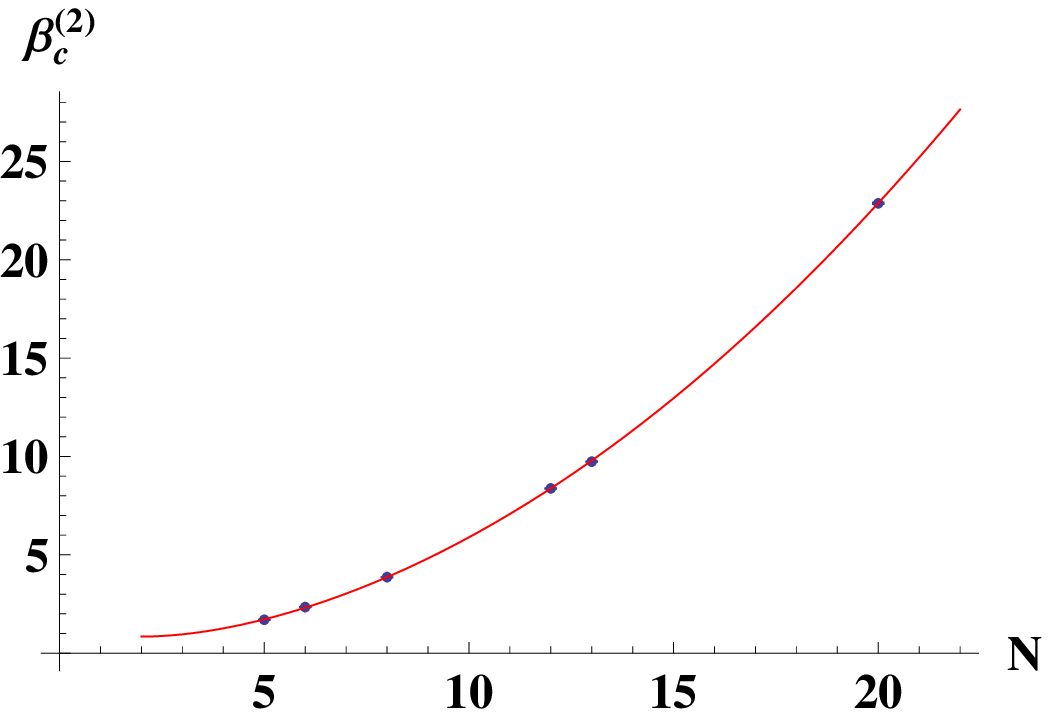}
\includegraphics[width=0.32\textwidth]{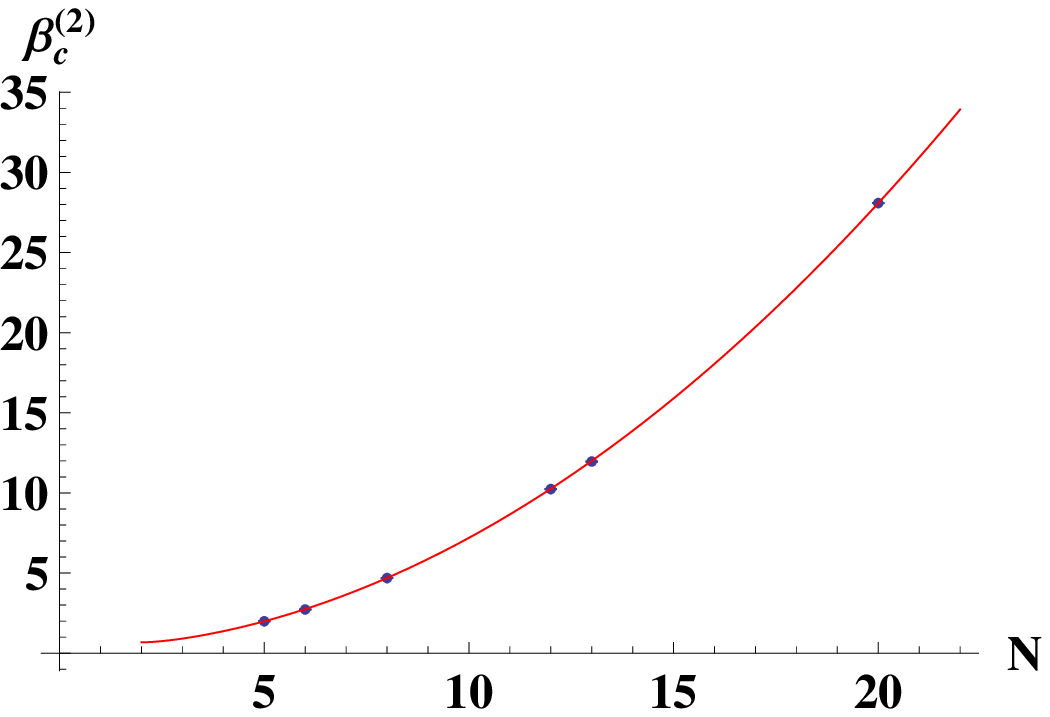}
\includegraphics[width=0.32\textwidth]{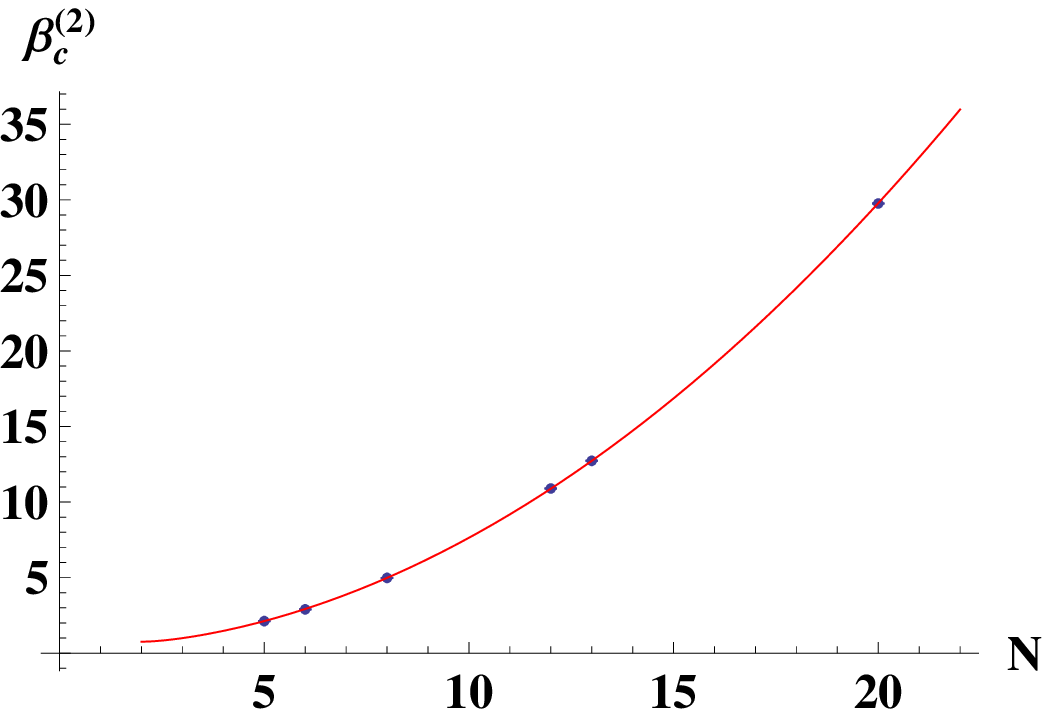}
\caption{Scaling function $A/(1-\cos{2\pi/N}) + B (1-\cos{2\pi/N})$
{\it versus} data for $\beta_{\rm c}^{(2)}$ at $N_t=2,\ 4,\ 8$ (from left to 
right).}
\label{fig:ndep2}
\end{figure}

For the critical couplings at the second transition, $\beta_{\rm c}^{(2)}$,
where determinations for many values of $N$ are available, we tried to
find a simple scaling dependence with $N$ at fixed $N_t$. Taking inspiration 
from Ref.~\cite{bhanot}, we made a few attempts and found that a good fit is 
achieved with the function
\[
\beta_{\rm c}^{(2)}(N) = \frac{A}{(1-\cos{2\pi/N})} + B (1-\cos{2\pi/N})\;.
\]
In Table~\ref{tbl:ndep2} we report the values of the parameters $A$ and $B$
for $N_t=2,\ 4,\ 8$, while Figs.~\ref{fig:ndep2} shows the fitting functions 
against numerical data. 

\begin{figure}[tb]
\centering
\includegraphics[width=0.32\textwidth]{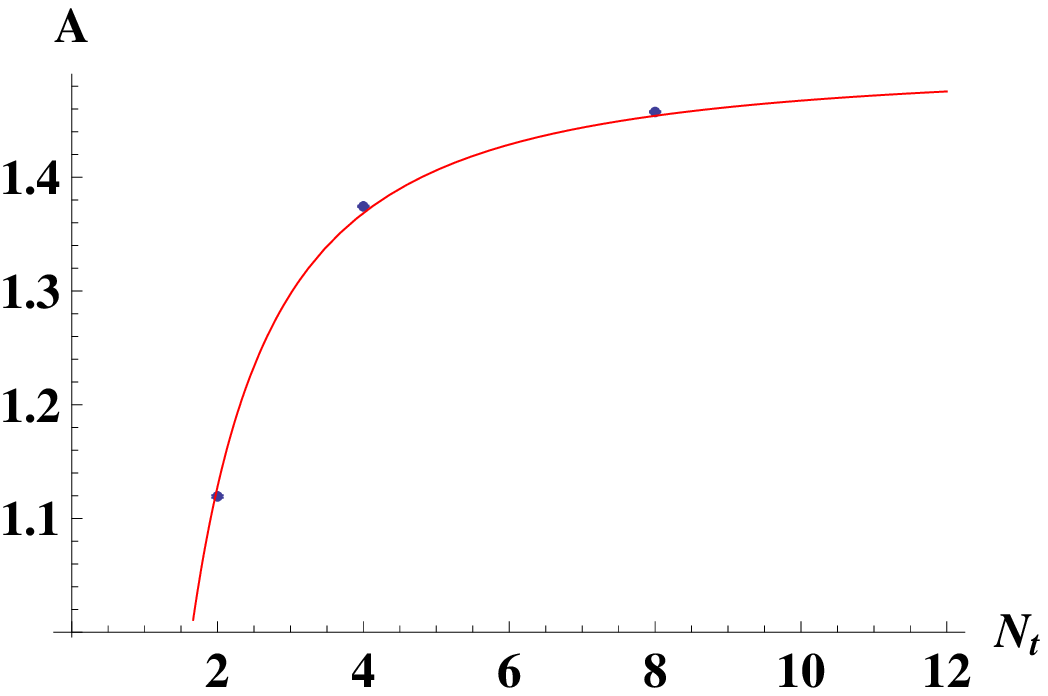}
\includegraphics[width=0.32\textwidth]{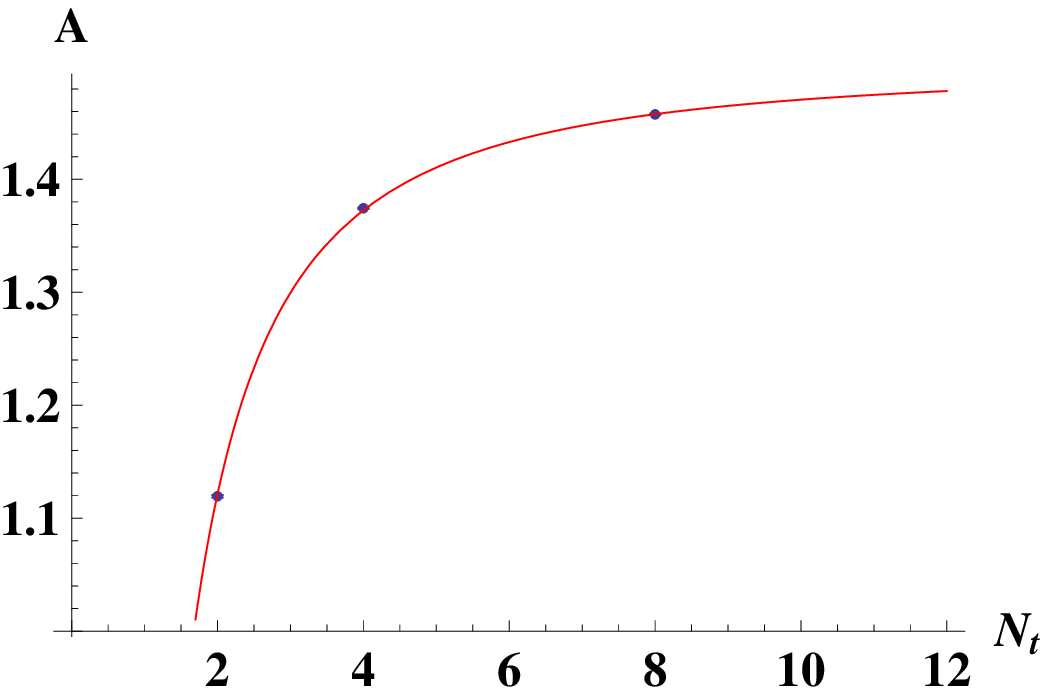}
\includegraphics[width=0.32\textwidth]{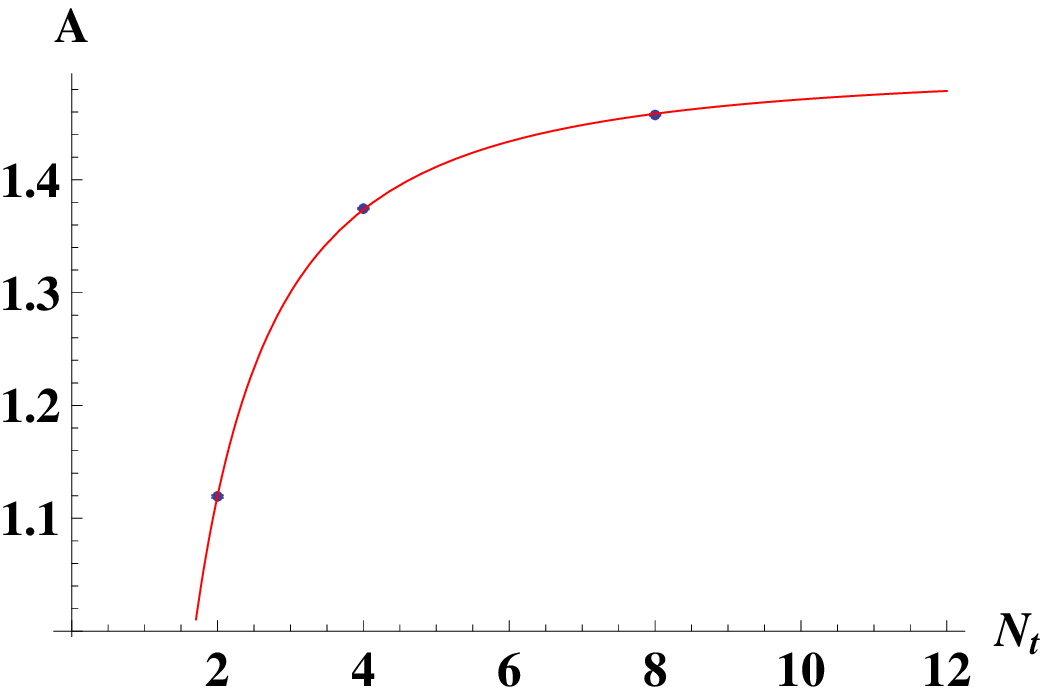}
\caption{Scaling for coefficient $A$ versus value of $N_t$ 
with different $\nu$ values ($\nu = 0.67,\ 0.64,\ 0.6331$ from left to right).}
\label{fig:ntadep}
\end{figure}

Inspecting the behavior of the coefficient $A$ in Table~\ref{tbl:ndep2}, 
one observes that it approaches its zero-temperature limit 
$A_{\infty}=1.50122$ when $N_t$ increases~\cite{ZN_zero_T}. 
We have investigated this approach in details and found that it can be well 
described by the following fitting function: 
$A = A_{\infty} - C N_t^{-1/\nu}$. The results of the fits are given in 
Table~\ref{tbl:ntadep} and Figs.~\ref{fig:ntadep}. It can be seen both from the 
$\chi^2_{\rm r}$ given in the table
 and from the plots that $\nu \approx 0.64$ describes data better than $\nu \approx 0.67$.

\begin{table}[tb]
\caption{Parameters of the scaling with $N$ of the second transition point, 
$\beta_{\rm c}^{(2)} = A /{(1-\cos{2\pi/N})} + B (1-\cos{2\pi/N})$ at fixed 
$N_t$.}
\label{tbl:ndep2}
\begin{center}
\begin{tabular}{||c||c|c|c||}
\hline
$N_t$ & $A$      & $B$         & $\chi^2_{\rm r}$ \\
\hline                                        
 2 & 1.1194(11)  & 0.141(24)   & 209  \\
 4 & 1.37440(60) & -0.0046(88) & 18.2 \\
 8 & 1.45745(57) & 0.0155(53)  & 16.1 \\
\hline
\end{tabular}
\end{center}
\end{table}

\begin{table}[tb]
\caption{Parameters of the fit of the $A$ dependence on $N_t$ with the 
scaling function $A = A_{\infty} - C N_t^{-1/\nu}$, with 
$A_{\infty} = 1.50122$, taken from Ref.~\cite{ZN_zero_T}; $\nu$ 
values given without errors are fixed at the known results of the 
zero-temperature theory~\cite{ZN_zero_T}.}
\label{tbl:ntadep}
\begin{center}
\begin{tabular}{||c|c|c||}
\hline
$C$         & $\nu$     & $\chi^2_{\rm r}$ \\
\hline                                            
 1.050(25)  & 0.67       & 95.0    \\
 1.1220(67) & 0.64       & 6.12    \\
 1.140(17)  & 0.6331(61) & 5.42    \\
\hline
\end{tabular}
\end{center}
\end{table}

\subsection{Continuum limit}

Finding the continuum limit of the finite temperature theory in the first
or in the second transition amounts to extrapolate the corresponding critical
couplings, $\beta_{\rm c}^{(1)}$ or $\beta_{\rm c}^{(2)}$, to the limit
$N_t\to\infty$ at fixed $N$.

The theory of dimensional cross-over~\cite{caselle} suggests the fitting
function to be used:
\begin{equation}
\beta_{\rm c}^{(1,2)}(N_t) = \beta_{{\rm c},\ T=0}^{(1,2)}
- (N_t a T_{\rm c})^{-1/\nu}\;,
\label{fit_cont}
\end{equation}
where $\beta_{{\rm c},\ T=0}^{(1,2)}$ and $\nu$ are the critical couplings 
and the critical index of the zero-temperature theory. Since we know that,
for any $N$, the $3D$ $Z(N)$ LGT at zero-temperature exhibits only one
phase transition, with the critical index $\nu$ depending on the side from
which the transition is approached~\cite{ZN_zero_T}, we expect that, for
a given $N$, the fit parameters $\beta_{{\rm c},\ T=0}^{(1)}$ and 
$\beta_{{\rm c},\ T=0}^{(2)}$ take the same value and agree with the 
zero-temperature critical coupling at the same $N$. 
As for the fit parameter $\nu$, we expect
it to agree with the value of the critical index $\nu$ at one of the two
sides of the zero-temperature transition.

We fitted with the function given in~(\ref{fit_cont}) our data for the
critical couplings $\beta_{\rm c}^{(2)}(N_t)$ at $N$=5,6,8,12,13,20
(see Tables~\ref{tbl:clim2} and~\ref{tbl:clim2bis}) and for the
critical couplings $\beta_{\rm c}^{(1)}(N_t)$ at $N$=5
(see Table~\ref{tbl:clim1}). In some cases in the fit we fixed 
either $\beta_{{\rm c},\ T=0}^{(1,2)}$ or $\nu$, or both, at the values known
from the zero-temperature theory~\cite{ZN_zero_T}. The scenario which
emerges from the inspection of Tables~\ref{tbl:clim2}, \ref{tbl:clim2bis}
and~\ref{tbl:clim1} is that, despite the large reduced chi-squared obtained
in a few cases, the agreement between the fit parameters 
$\beta_{{\rm c},\ T=0}^{(1,2)}$ and the known zero-temperature critical 
couplings~\cite{ZN_zero_T} is satisfactory. As for the value of the fit
parameter $\nu$, results are not precise enough to discriminate
between the known values of the critical index $\nu$ of the zero-temperature 
theory at one or the other side of the transition~\cite{ZN_zero_T}.

This analysis allows us for the determination of the critical temperature 
$a T_{\rm c}$ in the continuum limit for all the values of $N$ considered
in this work.

\begin{table}[tb]
\caption{Results of the fit of $\beta_{\rm c}^{(2)}(N_t)$ for $N=5,\ 6,\ 8$ 
with the function given in~(\ref{fit_cont}). Parameters are given without 
errors when their values were fixed at the known results of the 
zero-temperature corresponding theory~\cite{ZN_zero_T} (for the $\nu$ index 
we considered
both the values at the left and at the right of the zero-temperature
critical point). Parameters are given with a (-) mark when their errors 
are unavailable and with a ${}^{*}$ mark when obtained from fits on data 
with $N_t=4,\ 8$ only (in general, $N_t=2,\ 4,\ 8$ were considered).}
\label{tbl:clim2}
\begin{center}
\begin{tabular}{|c|c|c|c|c|}
\hline
$N$ & $aT_{\rm c}$ & $\beta_{{\rm c},\ T=0}^{(2)}$ & $\nu$ & $\chi_{\rm r}^2$\\
\hline
    & 0.868(-)          & 2.23055(-) & 0.877(-)  & -            \\
    & 0.813(27)         & 2.177(12)  & 0.670     & 158          \\
    & 0.803(30)         & 2.170(14)  & 0.640     & 223          \\ 
5   & 0.825(38)         & 2.17961    & 0.692(45) & 131          \\
    & 0.810(13)         & 2.17961    & 0.670     & 81.8         \\
    & 0.776(31)${}^{*}$ & 2.17961    & 0.670     & 74.2${}^{*}$ \\
    & 0.789(17)         & 2.17961    & 0.640     & 161          \\ 
    & 0.731(18)${}^{*}$ & 2.17961    & 0.640     & 31.4${}^{*}$ \\ 
\hline                      
    & 0.6814(-)         & 3.04317(-) & 0.876(-)  & -            \\
    & 0.6769(76)        & 2.977(10)  & 0.674     & 5.02         \\
    & 0.6740(85)        & 2.969(12)  & 0.642     & 6.90         \\ 
6   & 0.6832(46)        & 3.00683    & 0.768(15) & 1.14         \\
    & 0.6573(47)        & 3.00683    & 0.674     & 22.6         \\
    & 0.572(13)${}^{*}$ & 3.00683    & 0.674     & 1.44${}^{*}$ \\
    & 0.6487(60)        & 3.00683    & 0.642     & 40.6         \\ 
    & 0.542(21)${}^{*}$ & 3.00683    & 0.642     & 4.48${}^{*}$ \\ 
\hline
    & 0.42330(-)        & 5.14422(-) & 0.674(-)  & -            \\
    & 0.42378(12)       & 5.14299(25)& 0.672     & 0.19         \\
    & 0.4316(22)        & 5.1225(46) & 0.637     & 66.5         \\ 
8   & 0.4294(12)        & 5.12829    & 0.648(6)  & 33.0         \\
    & 0.4287(39)        & 5.12829    & 0.672     & 321          \\
    & 0.4427(39)${}^{*}$& 5.12829    & 0.672     & 177${}^{*}$  \\
    & 0.4298(19)        & 5.12829    & 0.637     & 86.1         \\ 
    & 0.4216(10)${}^{*}$& 5.12829    & 0.637     & 2.21${}^{*}$ \\ 
\hline     
\end{tabular}
\end{center}
\end{table}   
              
\begin{table}[tb]
\caption{Same as Table~\ref{tbl:clim2} for $N=12,\ 13,\ 20$.}
\label{tbl:clim2bis}
\begin{center}
\begin{tabular}{|c|c|c|c|c|}
\hline
$N$ &$aT_{\rm c}$ & $\beta_{{\rm c},\ T=0}^{(2)}$ & $\nu$ & $\chi_{\rm r}^2$ \\
\hline
    & 0.24728(-)         & 11.2566(-)  & 0.674(-)  & -            \\
    & 0.24559(13)        & 11.2640(23) & 0.670     & 0.22         \\
    & 0.25615(72)        & 11.218(12)  & 0.640     & 6.18         \\ 
12  & 0.2602(32)         & 11.1962     & 0.630(11) & 14.2         \\
    & 0.24954(28)        & 11.1962     & 0.670     & 89.8         \\
    & 0.2619(87)${}^{*}$ & 11.1962     & 0.670     & 55.5${}^{*}$ \\
    & 0.25742(10)        & 11.1962     & 0.640     & 12.7         \\ 
    & 0.2597(51)${}^{*}$ & 11.1962     & 0.640     & 21.3${}^{*}$ \\ 
\hline
    & 0.22433(-)         & 13.1391(-)  & 0.654(-)  & -            \\
    & 0.21872(53)        & 13.1656(56) & 0.671     & 5.88         \\
    & 0.22851(40)        & 13.1199(42) & 0.642     & 3.40         \\ 
13  & 0.2310(12)         & 13.1077     & 0.635(4)  & 8.86         \\
    & 0.2225(30)         & 13.1077     & 0.671     & 314          \\
    & 0.2342(62)${}^{*}$ & 13.1077     & 0.671     & 113${}^{*}$  \\
    & 0.22928(67)        & 13.1077     & 0.642     & 16.0         \\ 
    & 0.2311(24)${}^{*}$ & 13.1077     & 0.642     & 19.2${}^{*}$ \\ 
\hline
    & 0.144857(-)        & 30.5427(-)  & 0.608(-)  & -            \\
    & 0.1297(37)         & 30.73(10)   & 0.673     & 147          \\
    & 0.1356(24)         & 30.658(64)  & 0.647     & 58.8         \\ 
20  & 0.1357(26)         & 30.6729     & 0.642(19) & 58.2         \\
    & 0.13171(98)        & 30.6729     & 0.673     & 97.3         \\
    & 0.13199(13)${}^{*}$& 30.6729     & 0.673     & 1.57${}^{*}$ \\
    & 0.13506(54)        & 30.6729     & 0.647     & 31.0         \\ 
    & 0.13519(49)${}^{*}$& 30.6729     & 0.647     & 23.9${}^{*}$ \\ 
\hline
\end{tabular}
\end{center}
\end{table}

\begin{table}[tb]
\caption{Same as Table~\ref{tbl:clim2} for $\beta_{\rm c}^{(1)}(N_t)$ for 
$N$=5.}
\label{tbl:clim1}
\begin{center}
\begin{tabular}{|c|c|c|c|c|}
\hline
$N$ & $aT_{\rm c}$ & $\beta_{{\rm c},\ T=0}^{(1)}$ & $\nu$ & $\chi_{\rm r}^2$\\
\hline
    & 0.790(5)  & 2.198(9)  & 0.84(3)   & 1.21  \\
    & 0.764(14) & 2.144(9)  & 0.670     & 23.1  \\
    & 0.758(16) & 2.135(11) & 0.640     & 33.6  \\ 
5   & 0.786(7)  & 2.17961   & 0.788(10) & 2.66  \\
    & 0.722(16) & 2.17961   & 0.670     & 105   \\
    & 0.709(19) & 2.17961   & 0.640     & 171   \\ 
\hline     
\end{tabular}
\end{center}
\end{table}   

\subsection{Critical indices}

Some critical indices at the two transitions in the $3D$ $Z(N)$ LGT at
finite temperature can be extracted by the standard FSS analysis.
In particular, the behavior on the lattice size $L$ of the standard
magnetization $M_L$ and of its susceptibility at the second transition
allows to extract the indices $\beta/\nu$ and $\gamma/\nu$ through a
fit with the functions
\begin{eqnarray}
M_L &=& A_{M_L} L^{-\beta/\nu} \ \, \nonumber \\
\chi_{M_L} &=& A_{\chi_{M_L}} L^{\gamma/\nu} \ .
\label{Ldep}
\end{eqnarray} 
Similarly, the behavior on $L$ of the rotated magnetization $M_R$ and of
its susceptibility at the first transition point allow the extraction 
of the same critical indices at that transition.

Thereafter, the hyperscaling relation $2 \beta/\nu + \gamma/\nu = 2$ can be
checked and the magnetic index $\eta= 2 - \gamma/\nu$ can be extracted
at both transitions.

Our results are summarized in Tables~\ref{tbl:indices_first} 
and~\ref{tbl:indices_second}. We can see that the hyperscaling relation
is generally satisfied and the critical index $\eta$ generally takes values
compatible with 1/4 at the second transition and with $4/N^2$ at the first
transition, in agreement with the expectations.

\begin{table}[tb]
\caption{Critical indices at the first transition point obtained in $3D$ 
$Z(N)$ with $N=5,\ 8,\ 13$ and 20 for various values of $N_t$.}  
\label{tbl:indices_first}
\begin{center}
\begin{tabular}{|c|c|c|c|c|c|c|c|c|}
\hline
$N$ & $N_t$ & $\beta_{\rm c}^{(2)}$ & $\beta/\nu$ & $\chi_{\rm r}^2$ 
    & $\gamma/\nu$ & $\chi_{\rm r}^2$ & $d$ & $\eta$ \\
\hline
5 & 2 & 1.617(2) & 0.097(6)  & 0.101& 1.847(5)  & 0.561& 2.04(2)  & 0.153(5) \\
5 & 4 & 1.943(2) & 0.11(1)   & 1.25 & 1.841(1)  & 0.70 & 2.07(3)  & 0.159(1) \\
5 & 8 & 2.085(2) & 0.09(2)   & 0.77 & 1.844(1)  & 0.78 & 2.01(4)  & 0.156(1) \\
\hline
8 & 4 & 2.544(8) &$-$0.26(2) & 1.79 &  1.930(3) & 1.58 & 1.41(5)  & 0.070(3) \\
8 & 8 & 3.422(9) &$-$0.52(5) & 0.21 &  1.959(1) & 0.21 & 0.9(1)   & 0.040(1) \\
\hline
13 & 2 & 1.795(4)& 0.07(5)   & 1.28 & 1.968(9)  & 0.97 & 2.1(1)   & 0.032(9) \\
13 & 4 & 2.74(5) &$-$0.26(2) & 1.81 & 1.976(3)  & 1.80 & 1.5(1)   & 0.024(3) \\
13 & 8 & 3.358(7)&$-$0.9(1)  & 1.17 & 1.973(4)  & 1.25 & 0.3(2)   & 0.027(4) \\
\hline
20 & 4 & 2.57(1) &$-$0.25(2) & 0.37 &  1.991(3) & 1.91 &  1.49(5) & 0.009(3) \\
20 & 8 & 3.42(5) &$-$0.72(6) & 0.41 & 1.9790(16)& 0.33 & 0.55(13) & 0.0210(16)
\\
\hline
\end{tabular}
\end{center}
\end{table}

\begin{table}[tb]
\caption{Critical indices at the second transition point obtained in $3D$ 
$Z(N)$ with $N = 5,\ 6,\ 8,\ 12,\ 13$ and 20 for various values of $N_t$.}  
\begin{center}
\begin{tabular}{|c|c|c|c|c|c|c|c|c|}
\hline
$N$ & $N_t$ & $\beta_{\rm c}^{(2)}$ & $\beta/\nu$ & $\chi_{\rm r}^2$ 
    & $\gamma/\nu$ & $\chi_{\rm r}^2$ & $d$ & $\eta$ \\
\hline
5 & 2 & 1.6972(14)& 0.1259(2) & 1.22 & 1.750(3)	& 0.50 & 2.001(4) & 0.250(3)\\
5 & 4 & 1.9885(15)& 0.1061(3) &	2.67 & 1.758(9)	& 2.45 & 1.971(9) & 0.242(9)\\
5 & 8 & 2.1207(9) & 0.1376(6) & 2.04 & 1.747(15)& 1.62 & 2.022(16)& 0.253(15)\\
\hline
6 & 2 & 2.3410(15)& 0.26(3)   & 1.8  & 1.6(6)   & 1.21 & 2.1(6)	  & 0.4(6) \\
6 & 4 & 2.725(12) & 0.1056(13)&	1.84 & 1.76(7)	& 2.05 &1.97(8)   & 0.24(7)\\ 
6 & 8 & 2.899(4)  & 0.0949(4) & 1.67 & 1.731(8)	& 0.71 & 1.920(9) & 0.269(8)\\ 
\hline
8 & 2 & 3.8640(10)& 0.1336(4) & 0.36 & 1.743(15)& 0.73 & 2.010(16)& 0.257(15)\\
8 & 4 & 4.6864(15)& 0.1278(4) & 3.85 & 1.753(6)	& 1.34 & 2.009(7) & 0.247(6) \\
8 & 8 & 4.9808(5) & 0.1379(5) & 0.77 & 1.745(18)& 1.82 & 2.020(19)& 0.255(18)\\
\hline
12 & 2 & 8.3745(5)& 0.1283(16)& 1.22 & 1.73(4)  & 0.78 & 1.99(4)  & 0.27(4) \\ 
12 & 4 & 10.240(7)& 0.1303(4) & 1.52 & 1.746(9) & 0.87 & 2.007(10)& 0.254(9)\\ 
12 & 8 & 10.898(5)& 0.149(3)  & 0.64 & 1.78(16)	& 1.19 & 2.07(17) & 0.22(16)\\ 
\hline
13 & 2 & 9.735(4) & 0.1251(2) & 0.22 & 1.744(5)	& 0.09 & 1.995(5) & 0.256(5)\\
13 & 4 & 11.959(6)& 0.1265(2) & 1.43 & 1.746(3) & 0.48 & 1.999(4) & 0.254(3)\\ 
13 & 8 & 12.730(2)& 0.1357(18)& 3.55 & 1.75(2)	& 0.82 & 2.02(2)  & 0.25(2) \\ 
\hline
20 & 2 & 22.87(4) & 0.1322(14)& 1.06 & 1.78(3)  & 0.68 & 2.04(4)  & 0.22(3) \\ 
20 & 4 & 28.089(3)& 0.1384(4) & 0.17 & 1.748(14)& 0.17 & 2.025(15)& 0.252(14)\\
20 & 8 & 29.758(6)& 0.1278(7) & 1.60 & 1.713(17)& 1.15 & 1.968(18)& 0.287(17)\\
\hline
\end{tabular}
\end{center}
\label{tbl:indices_second}
\end{table}

Finally, the critical index $\nu$ at the second transition was estimated 
following a procedure inspired by Ref.~\cite{3dxy_univ}: first, for each 
lattice size $L$ the known function $U_L^{(M)}(\beta)$ is used to determine 
$d U_L^{(M)}(\beta)/d\beta$; then, from this, the derivative of $U_L^{(M)}$ 
with respect to the rescaled coupling $x = (\beta - \beta^{(2)}_{\rm c}) 
(\ln L)^{1/\nu}$ can be calculated,
\begin{equation}
\frac{d U_L^{(M)}} {d x} = \frac{d U_L^{(M)}}{d \beta} (\ln L)^{1/\nu}\;.
\label{nu_determ}
\end{equation}
The best estimate of $\nu$ is found by minimizing the deviation of 
$d U_L^{(M)}/d x$ with respect to a constant value. The minimization was
done at $\beta^{(2)}_{\rm c}$. The resulting values for $\nu$, summarized in 
Table~\ref{tbl:nu_index}, exhibit a fair agreement with the expected BKT value
1/2, sometimes within large error bars.

\begin{table}[tb]
\caption{Critical index $\nu$ at the second transition point
in $3D$ $Z(N)$ with $N = 5,\ 6,\ 8,\ 12,\ 13$ and 20 for various values of 
$N_t$.}  
\label{tbl:nu_index}
\begin{center}
\begin{tabular}{|c|c|c|c|}
\hline
$N$ & $N_t$ & $\nu$ & $\chi^2$  \\
\hline
 5  & 2 & 0.55(9)  & 1.06 \\
 5  & 4 & $2\pm 5$ & 7.89 \\ 
 5  & 8 & 0.46(4)  & 0.51 \\ 
\hline
 6  & 2 & -        & -    \\ %
 6  & 4 & 0.5(2)   & 0.42 \\ %
 6  & 8 & 0.57(10) & 0.20 \\ %
\hline
 8  & 2 & 0.63(5)  & 0.16 \\ %
 8  & 4 & 0.52(16) & 1.01 \\ 
 8  & 8 & 0.42(3)  & 1.32 \\ 
\hline
12  & 2 & 0.41(13) & 0.018\\ %
12  & 4 & 0.60(8)  & 0.20 \\ %
12  & 8 & 0.33(1)  & 0.008\\ %
\hline
13  & 2 & $1\pm 2$ & 3.27 \\ 
13  & 4 & 0.62(9)  & 0.34 \\ 
13  & 8 & 0.43(3)  & 0.83 \\ 
\hline
20  & 2 & 0.60(12) & 0.39 \\ %
20  & 4 & 0.57(4)  &  0.05\\ 
20  & 8 & 0.39(2)  & 0.13 \\ 
\hline
\end{tabular}
\end{center}
\end{table}

\section{Summary} 

This paper completes our study of the critical behavior of $3D$ $Z(N>4)$ 
lattice gauge theories both at zero and at finite temperatures. In order to 
accomplish this investigation, we have used various methods like 
renormalization group, duality transformations and Monte Carlo 
simulations, combined with finite-size scaling. Here we would like to outline 
our main findings and list some open problems left for future investigation. 

The main results can be shortly summarized as follows. 

\begin{itemize} 
\item We have found that in all $Z(N)$ vector models two BKT-like phase 
transitions occur at finite temperatures if $N>4$. In this paper we have 
extended the results of Ref.~\cite{ZN_fin_T} for $N=5,\ 13$ and $N_t=2,\ 4$ to 
other values of $N$ and to $N_t=8$. In all cases studied, the results for the 
critical indices suggest that finite-temperature $Z(N)$ lattice models belong 
to the universality class of two-dimensional $Z(N)$ vector spin models, in 
agreement with the Svetitsky-Yaffe conjecture. Furthermore, the available 
results for many values of $N$ allowed us to propose and check some
scaling formulas for the critical point of the second phase transition. 
Combining the results of the present paper with those for the index $\nu$ 
obtained by us at zero temperature in Ref.~\cite{ZN_zero_T} enabled us to 
check the continuum scaling and to predict the approximate value for $a T_c$ 
in the continuum limit. 

\item Three-dimensional $Z(N>4)$ models at zero temperature exhibit a single 
phase transition which appears to be of 3rd order if one approaches the 
critical point from above and belongs to the universality class of the 
$3D$ $XY$ model. However, if one approaches the critical point from below, 
the index $\alpha$ is compatible with the value of the $3D$ Ising model. This 
suggests that the free energy develops a cusp in the large volume limit which 
leads to different singularities. 
A very interesting feature of all $Z(N)$ models at large $N$ is the existence 
of a $U(1)$ symmetric phase on the finite lattice which manifests itself in 
the characteristic behavior of the scatter plots for magnetization 
of the dual spins. 

\end{itemize} 
 
The most interesting open problems, on our opinion, are the following. 
 
\begin{itemize} 
\item  More precise determination of critical points and indices at the
1st BKT transition for $N\geq 8$ and establishing the formula for the 
scaling of these critical points with $N$. 
 
\item What is the physics behind the two BKT transitions? Most probably it is 
related to the existence of two topological excitations dual to each other. 
This would be similar to what happens in $2D$ $Z(N)$ spin models. 
Unfortunately, the analytical proof of this is still to be constructed. 
 
\item A very intriguing problem is the physics of symmetric phase at zero 
temperature. In particular, it is not clear if this phase survives the 
transition to the thermodynamical limit and how it is connected to the 
massless BKT phase at finite temperature.
 
\end{itemize}

\section{Acknowledgments} 

The work of the Ukrainian co-authors was supported by the Ukrainian State Fund 
for Fundamental Researches under the grant F58/384-2013. Numerical simulations
have been partly carried out on Ukrainian National GRID facilities.
O.B. thanks for warm hospitality the Dipartimento di Fisica dell'Universit\`a 
della Calabria and the INFN Gruppo Collegato di Cosenza during the final 
stages of this investigation.
This work has been partially supported by the INFN SUMA project.

\end{document}